\documentclass[aps,prx,twocolumn,superscriptaddress,longbibliography,showkeys]{revtex4-1}

\usepackage{amsmath}
\usepackage{amsthm}
\usepackage{amssymb}
\usepackage{graphicx}
\usepackage{esint}
\usepackage{titlesec}
\usepackage{amsfonts}
\usepackage{xcolor}
\usepackage{times}

\usepackage{bbm}

\newcommand{\ket}[1]{\vert #1 \rangle}
\newcommand{\bra} [1] {\langle #1 \vert}

\newcommand{\ad}{\hat{a}^{\dagger}}
\newcommand{\bd}{\hat{b}^{\dagger}}
\renewcommand{\a}{\hat{a}}
\renewcommand{\b}{\hat{b}}

\newcommand{\bs}{U^{\mathrm{BS}}_{\eta}}
\newcommand{\bsd}{U^{\mathrm{BS} \dagger}_{\eta}}
\newcommand{\tms}{U^{\mathrm{PDC}}_{g}}
\newcommand{\tmsd}{U^{\mathrm{PDC} \dagger}_{g}}

\begin{document}

\title{Two-boson quantum interference in time}

\author{Nicolas J. Cerf}
\affiliation{Centre for Quantum Information and Communication, Ecole polytechnique de Bruxelles, CP 165/59, Universit\'e libre de Bruxelles, 1050 Brussels, Belgium}

\author{Michael G. Jabbour}
\affiliation{Department of Applied Mathematics and Theoretical Physics, Centre for Mathematical Sciences, University of Cambridge, Cambridge CB3 0WA, United Kingdom}


\keywords{Quantum interference $|$ boson bunching $|$ time reversal} 

\begin{abstract}
\textbf{The celebrated Hong-Ou-Mandel effect is the paradigm of two-particle quantum interference. It has its roots in the symmetry of identical quantum particles, as dictated by the Pauli principle. Two identical bosons impinging on a beam splitter (of transmittance~1/2) cannot be detected in coincidence at both output ports, as confirmed in numerous experiments with light or even matter. Here, we establish that \textit{partial time reversal} transforms the beamsplitter linear coupling into amplification. We infer from this duality the existence of an unsuspected two-boson interferometric effect in a quantum amplifier (of gain~2) and identify the underlying mechanism as \textit{timelike indistinguishability}. This fundamental mechanism is generic to any bosonic Bogoliubov transformation, so we anticipate wide implications in quantum physics.}
\end{abstract}

\maketitle

\begin{figure}[t]
\centering
\includegraphics[trim = 0cm 0cm 0cm 0cm, clip, width=0.8\linewidth, page=1]{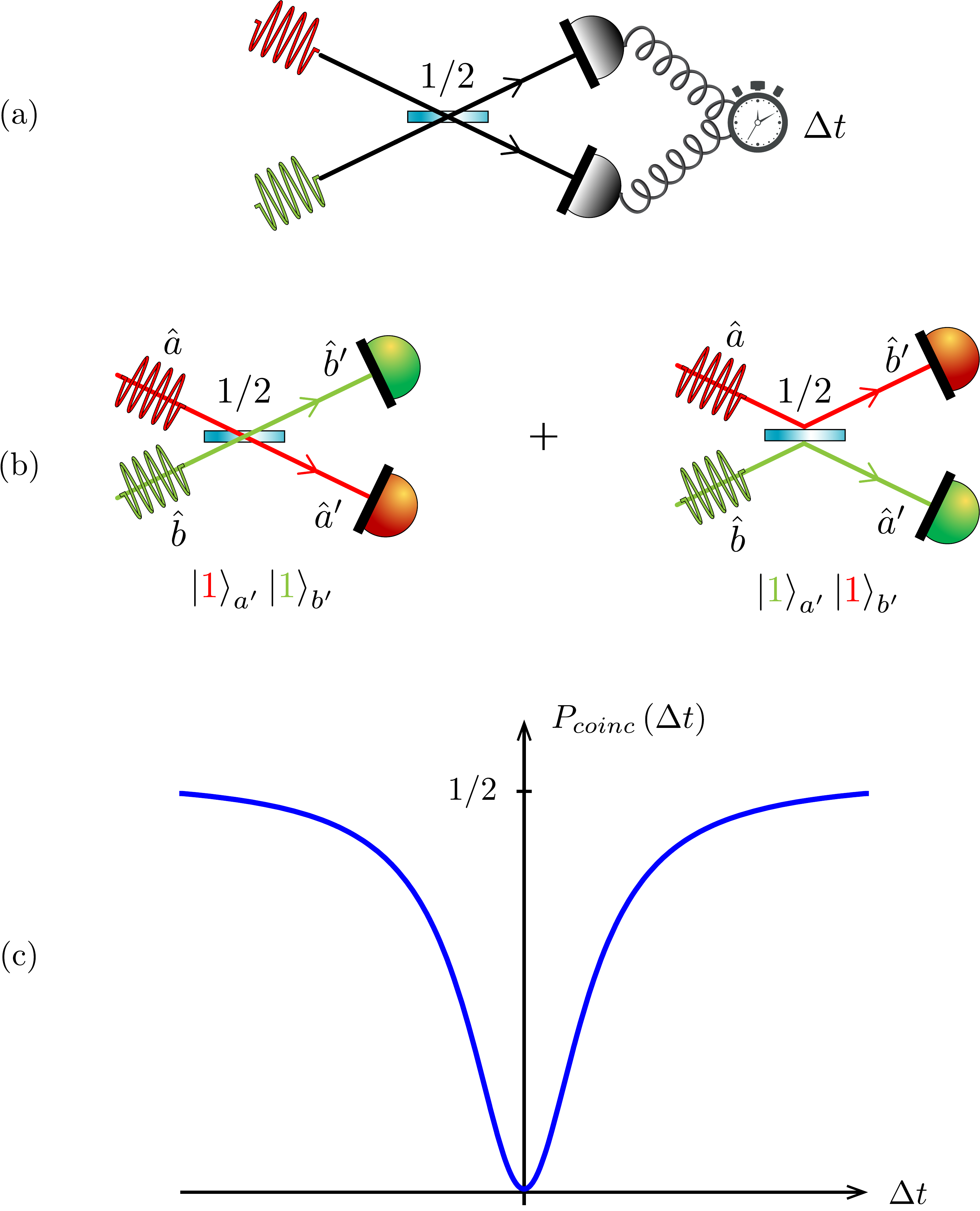}
\caption{ (a) If two indistinguishable photons (represented in red and green for the sake of argument) simultaneously enter the two input ports of a 50:50 beam-splitter, they always exit the same output port together (no coincident detection can be observed). 
(b)  The probability amplitudes for double-transmission (left) and double-reflection (right) precisely cancel each other when the transmittance is equal to 1/2. This is a genuinely quantum effect, which cannot be described as a classical wave interference.
(c) The correlation function exhibits a HOM dip when the time difference $\Delta t$ between the two detected photons is close to zero (i.e., when they tend to be indistinguishable).
}
\label{fig-HOM}
\end{figure}

The laws of quantum physics govern the behavior of identical particles via the symmetry of the wave function, as dictated by the Pauli principle \cite{Pauli}. In particular, it has been known since Bose and Einstein \cite{Einstein} that the symmetry of the bosonic wave function favors the so-called bunching of identical bosons. A striking demonstration of bosonic statistics for a pair of identical bosons was achieved in 1987 in a seminal experiment by Hong, Ou, and Mandel (HOM) \cite{HOM}, who observed the cancellation of coincident detections behind a 50:50 beam splitter (BS) when two indistinguishable photons impinge on its two input ports, see Fig.~\ref{fig-HOM}a. This HOM effect follows from the destructive two-photon interference between the probability amplitudes for double-transmission and double-reflection at the beam splitter, see Fig.~\ref{fig-HOM}b. Together with the Hanbury Brown and Twiss effect \cite{HBTwiss,Kimble} and the violation of Bell inequalities \cite{Bell,CHSH}, it is often viewed as the most prominent genuinely quantum feature: it highlights the singularity of two-particle quantum interference as it cannot be understood in terms of classical wave interference \cite{Pittman,Mandel}. It has been verified in numerous experiments over the last thirty years, see e.g. \cite{Yamamoto,Gisin,Tanzilli,Imoto}, even in case the single photons are simultaneously emitted by two independent sources \cite{Zeilinger,Grangier,Rarity} or within a silicon photonic chip \cite{Thompson,Bacco}. Remarkably, it has even been experimentally observed with $^4$He metastable atoms, demonstrating that this two-boson mechanism encompasses both light and matter \cite{Aspect}.
\par

\begin{figure}[t]
\centering
\includegraphics[trim = 0cm 0cm 0cm 0cm, clip, width=0.8\linewidth, page=2]{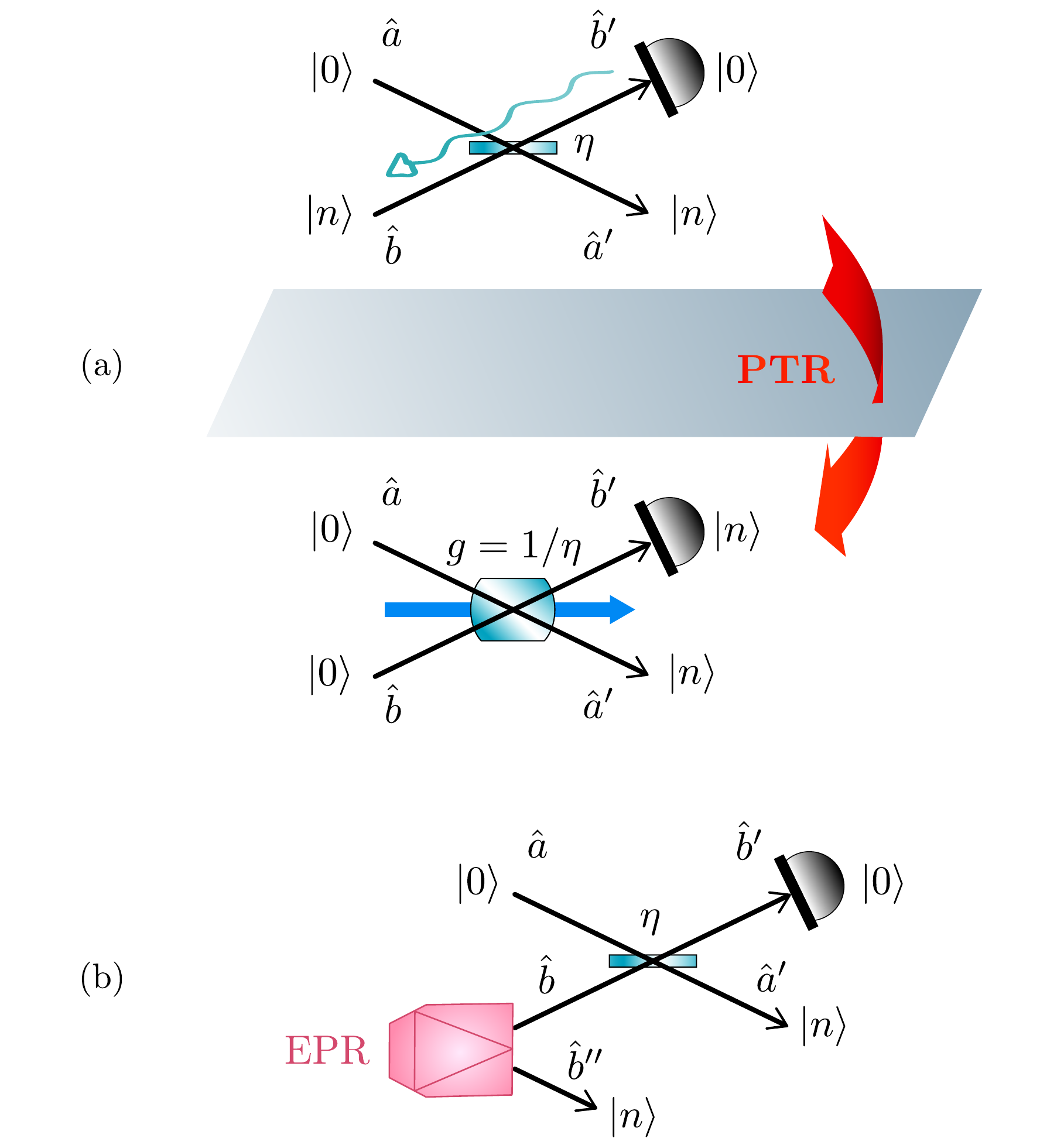}
\caption{(a) Beam splitter under partial time reversal (PTR), flipping the arrow of time in mode $\b$. The PTR duality is illustrated when $n$ photons impinge on port $\b$ (with vacuum on port $\a$) and we condition on all photons being reflected. The retrodicted state of mode $\b'$ (initially the vacuum state $\ket{0}$) backpropagates from the detector to the source (suggested by a wavy arrow). This yields the same transition probability amplitude (up to a constant) as for a PDC of gain $g=1/\eta$ with input state $\ket{0,0}$ and output state $\ket{n,n}$. PDC is an active Bogoliubov transformation, requiring a pump beam (represented in blue). Note that the PTR duality is rigorously valid when this pump beam is of high intensity (i.e., treated as a classical light beam) since the Hamiltonian $H_\mathrm{PDC}$ of Eq.~(\ref{eq-Hamiltonian-BS-PDC}) holds in this limit only.  
(b) Operational view of the PTR duality. As noted in  \cite{Levon}, if we prepare the entangled (EPR) state $\ket{\Psi}_{b,b''}\propto \sum_{n=0}^\infty \ket{n,n}$ and send mode $\b$ in the BS, we get the output state $\ket{\Psi}_{a',b''} \propto \sum_{n=0}^\infty \sin^n \!\theta \, \ket{n,n}$, which is precisely the two-mode squeezed vacuum state produced by PDC when the signal and idler modes are initially in the vacuum state. 
}
\label{fig-BS-under-PTR}
\end{figure}

Here, we explore how two-boson quantum interference transforms under reversal of the arrow of time in one of the two bosonic modes, see Fig.~\ref{fig-BS-under-PTR}a. This operation, which we dub partial time reversal (PTR), is unphysical but nevertheless central as it allows us to exhibit a duality between the linear optical coupling effected by a beam splitter and the nonlinear optical (Bogoliubov) transformation effected by a parametric amplifier. As a striking implication of these considerations, we predict a two-photon interferometric effect in a parametric amplifier of gain 2 (which is dual to a beam splitter of transmittance 1/2). We argue that this unsuspected effect originates from the indistinguishability between photons from the past and future, which we coin ``timelike'' indistinguishability as it is the partial time-reversed version of the usual ``spacelike'' indistinguishability that is at work in the HOM effect.
\par

Since Bogoliubov transformations are ubiquitous in quantum physics, it is expected that this two-boson interference effect in time could serve as a testbed for a wide range of bosonic transformations. Furthermore, from a deeper viewpoint, it would be fascinating to demonstrate the consequence of timelike indistinguishability in a photonic or atomic platform as it would help elucidating some heretofore overlooked fundamental property of identical quantum particles.
\par

\section*{Hong-Ou-Mandel effect}

The HOM effect is a landmark in quantum optics as it is the most spectacular manifestation of boson bunching. It is a two-photon intrinsically quantum interference effect where the probability amplitude of both photons being transmitted cancels out the probability amplitude of both photons being reflected. A 50:50 beam splitter effects the single-photon transformations (see Materials and Methods A)
\begin{equation}
|1\rangle_a \to \frac{1}{\sqrt{2}} ( \underbrace{|1\rangle_a}_\mathrm{trans} - \underbrace{|1\rangle_b}_\mathrm{ref}  ), \quad 
|1\rangle_b \to \frac{1}{\sqrt{2}}  ( \underbrace{|1\rangle_a}_\mathrm{ref} +  \underbrace{|1\rangle_b}_\mathrm{trans} ), 
\label{eq-single-photon-transformation}
\end{equation}
where $a$ and $b$ label the bosonic modes and $|1\rangle_{a/b}$ stands for a single-photon state in mode $a/b$ (here, ``trans'' stands for transmitted and ``ref'' for reflected).
Hence, the state of two indistinguishable photons (one in each mode) transforms as
\begin{equation}
|1\rangle_a |1\rangle_b \to \frac{1}{\sqrt{2}}  (|1\rangle_a |1\rangle_a  -  |1\rangle_b  |1\rangle_b), 
\label{eq-double-photon-transformation}
\end{equation}
since the double-transmission term $|1\rangle_a  |1\rangle_b$ cancels out the double-reflection term $|1\rangle_b |1\rangle_a$. 
More generally, the probability for coincident detections  is given by (see Materials and Methods E)
\begin{equation}
P_\mathrm{coinc} (\eta) = |\bra{1,1}  \bs \ket{ 1,1}|^2 = (2\eta-1)^2,
\label{eq-Pcoinc-BS}
\end{equation}
where $\bs$ denotes the unitary corresponding to a beam splitter of transmittance $\eta$. In a nutshell, the HOM effect boils down to
$\bra{1,1} U^{\mathrm{BS}}_{1/2} \ket{1,1} = 0$. 
Its experimental manifestation is the presence of a dip in the correlation function, witnessing that two photons cannot be coincidently detected at the two output ports when $\eta=1/2$, see Fig.~\ref{fig-HOM}c.
\par

\section*{Partial time reversal}

Bogoliubov transformations on two bosonic modes comprise passive and active transformations. The beam splitter (BS) is the fundamental passive transformation, while parametric down-conversion (PDC) gives rise to the class of active transformations (also called non-degenerate parametric amplification). Although the involved physics is quite different (a simple piece of glass makes a beam splitter, while an optically pumped nonlinear crystal is needed to effect PDC), the Hamiltonians generating these two unitaries are amazingly close, namely
\begin{equation}
H_\mathrm{BS} = i (\hat a^\dagger \hat b - \hat a \hat b^\dagger), \quad H_\mathrm{PDC} = i (\hat a^\dagger \hat b^\dagger - \hat a \hat b), 
\label{eq-Hamiltonian-BS-PDC}
\end{equation}
where $ \hat a$ and  $\hat b$ are mode operators.
It is striking that a simple swap $\hat b \leftrightarrow \hat b^\dagger$ transforms $H_\mathrm{BS} $ into $H_\mathrm{PDC}$, suggesting a deep duality between a beam splitter and a parametric down-converter by reversing the arrow of time in mode $\hat b$ (keeping mode $\hat a$ untouched). 

The underlying concept of partial time reversal (PTR) will be formalized in Eq. (\ref{eq-fundamental-equation}), but we first illustrate this duality between a BS and PDC with the simple example of Fig.~\ref{fig-BS-under-PTR}a, where $n$ photons impinge on port $\b$ of a BS (with vacuum on port $\a$), resulting in the binomial output state  
\begin{equation}
\bs \ket{0,n} = \sum_{k=0}^n \binom{n}{k}^{1/2} ( \sin\theta)^k (\cos\theta)^{n-k} \ket{k,n-k}  ,
\end{equation}
with $\eta = \cos^2 \theta$, see Materials and Methods A. The path where all photons are reflected ($k=n$) is associated with the transition probability amplitude $\bra{n,0} \bs \ket{0,n} =  \sin^n\theta$. Reversing the arrow of time on mode $\b$ (see Fig. \ref{fig-BS-under-PTR}a) leads us to consider the transition probability amplitude for a PDC of gain $g = \cosh^2 r$ with vacuum state on its two inputs and $n$ photon pairs on its outputs, that is, $\bra{n,n} \tms \ket{0,0} = \tanh^n r / \cosh r$, see Materials and Methods A. Strikingly, the above two amplitudes can be made equal (up to a constant $\cosh r$) if we set $\sin \theta = \tanh r$, or equivalently $\eta=1/g$. Thus, we have
\begin{equation}
\bra{n,n} \tms \ket{0,0} = g^{-1/2}  \bra{n,0} U^{\mathrm{BS}}_{1/g} \ket{0,n}   . 
\end{equation}
The case of $m$ photons (instead of vacuum) impinging on port $\a$ and $n$ photons impinging on port $\b$ is illustrated in Materials and Methods B. Conditioning again the output port $\hat b'$ on vacuum and reversing the arrow of time, we obtain the same output as for a PDC of gain $g=1/\eta$ with input state $\ket{m,0}$.

These examples reflect the existence of a general duality between a BS and PDC. Indeed, as demonstrated in Materials and Methods C, partial transposition in Fock basis gives rise to PTR duality
\begin{equation}
\bra{n,j} \tms \ket{i,m} = g^{-1/2} \bra{n,m} U^{\mathrm{BS}}_{1/g} \ket{i,j}   ,
\label{eq-fundamental-equation}
\end{equation}
where indices $j$ and $m$ have been swapped [this can alternatively be expressed as Eqs. (\ref{eq-fundamental-partial transpose}) or (\ref{eq-fundamental-equation-2})].
The PTR duality is nicely evidenced by the conservation rules exhibited by the BS and PDC transformations: the former conserves the total photon number, while the latter conserves the difference between the photon numbers. In Eq. (\ref{eq-fundamental-equation}), the only nonzero matrix elements for a BS are those satisfying $i+j=n+m$. This directly implies that the only nonzero matrix elements for a PDC satisfy $i-m=n-j$, as expected.
\par

The notion of time reversal can be conveniently interpreted using the so-called ``retrodictive'' picture of quantum mechanics \cite{Aharonov}. Along this line, PTR must be understood here as the fact that the ``retrodicted'' state of mode $\hat b$ propagates backwards in time, while the state of mode $\hat a$ normally propagates forwards in time (this is made precise in Materials and Methods D).  As shown in Fig.~\ref{fig-BS-under-PTR}b, the PTR duality can be made operational by sending half of a so-called Einstein-Podolsky-Rosen (EPR) entangled state on mode $\b$, so that we access the output retrodicted state on the second entangled mode $\hat b''$. 
\par

\begin{figure}[t]
\centering
\includegraphics[trim = 0cm 0cm 0cm 0cm, clip, width=0.7\linewidth]{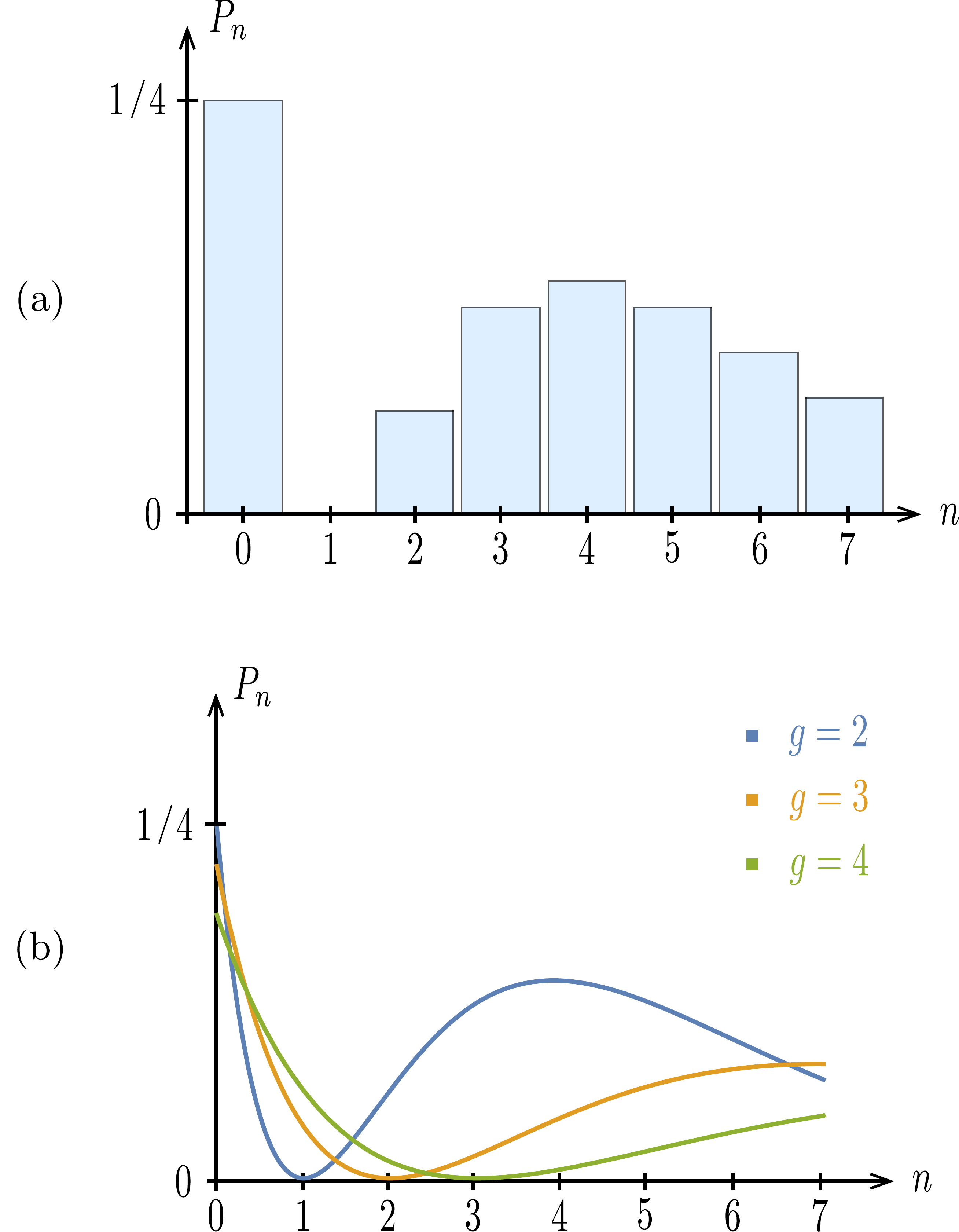}
\caption[]{(a) Probability $P_n$ of observing $n$ photon pairs at the output of a PDC of gain $g=2$ when a single photon impinges on both the signal and idler input ports. The output state\\  
\begin{minipage}{\linewidth}
\begin{eqnarray*}
\lefteqn{  U^{\mathrm{PDC}}_{2} \ket{1,1} = {\frac{1}{2}} \left(  - \ket{0,0} + \sqrt{\frac{1}{4}} \, \ket{2,2} + \sqrt{\frac{1}{2}} \, \ket{3,3}  \right. } \hspace{2.3 cm} \\
&&  \left. + \sqrt{\frac{9}{16}} \, \ket{4,4} + \sqrt{\frac{1}{2}} \, \ket{5,5} + \cdots \right) 
\end{eqnarray*}
\end{minipage}
has a vanishing $\ket{1,1}$ component, owing from two-photon interferometric suppression. The significant components are the vacuum as well as the terms with 2 to $\sim$10 pairs (the next terms quickly decay to zero). (b) Corresponding distributions of the photon pair number for a gain $g=3$ (showing a dip at $n=2$) and $g=4$ (showing a dip at $n=3$). The distributions are shown as continuous curves in order to guide the eye, but only integer values of $n$ are relevant. The curve for $g=2$ is also plotted for comparison.
}
\label{fig-output-state}
\end{figure}

\section*{Two-boson interference in an amplifier}

Due to this duality, the HOM effect for a BS of transmittance 1/2 immediately suggests the possible existence of a related interferometric suppression effect in a PDC of gain 2, namely $\bra{1,1} U^{\mathrm{PDC}}_{2} \ket{1,1} = 0$. This striking prediction can indeed be verified by examining the state at the output of a PDC of gain $g=2$ (see Materials and Methods E),
\begin{eqnarray}
\ket{\Psi}  \equiv U^{\mathrm{PDC}}_{2} \ket{1,1} = {\frac{1}{2}} \sum_{n=0}^\infty \frac{n-1}{2^{n/2}} \,  \ket{n,n}  .
\label{eq-distribution-n}
 \end{eqnarray}
As it appears, the component with a single photon on each output mode ($n=1$) is indeed suppressed, which is reminiscent of the HOM effect. However, the structure of $\ket{\Psi} $ is more complicated here as the PDC (unlike the BS) does not conserve the total photon number, hence we observe terms with $n\ge 2$ photons on each mode. The distribution of the photon-pair number is illustrated in Fig.~\ref{fig-output-state}a. 
 \par

The dependence of the probability of detecting a single pair ($n=1$) on the gain $g$ of PDC is given by (see Materials and Methods E)
\begin{equation}
P'_\mathrm{coinc} (g) = | \bra{1,1} \tms \ket{1,1} |^2 = (2-g)^2 / g^3  ,
\label{eq-Pcoinc-PDC}
\end{equation}
confirming that the probability for coincident detections vanishes if the gain $g=2$.
Note that if we substitute $\eta$ by $1/g$ in Eq.~(\ref{eq-Pcoinc-BS}) and divide by $g$, we get exactly Eq.~(\ref{eq-Pcoinc-PDC}), as implied by PTR duality.
More generally, we show in Materials and Methods F that this interferometric suppression effect actually extends to any larger integer value of the gain, e.g., $g=3,\, 4,\cdots$. As illustrated in Fig.~\ref{fig-output-state}b, the probability of detecting $n$ photons simultaneously on each output port vanishes when the gain $g=n+1$. Again, this is the consequence of PTR duality applied to an extended HOM effect.
\par

\section*{Spacelike vs. timelike indistinguishability}

\begin{figure}[t]
\centering
\includegraphics[trim = 0cm 0cm 0cm 0cm, clip, width=0.8\linewidth, page=4]{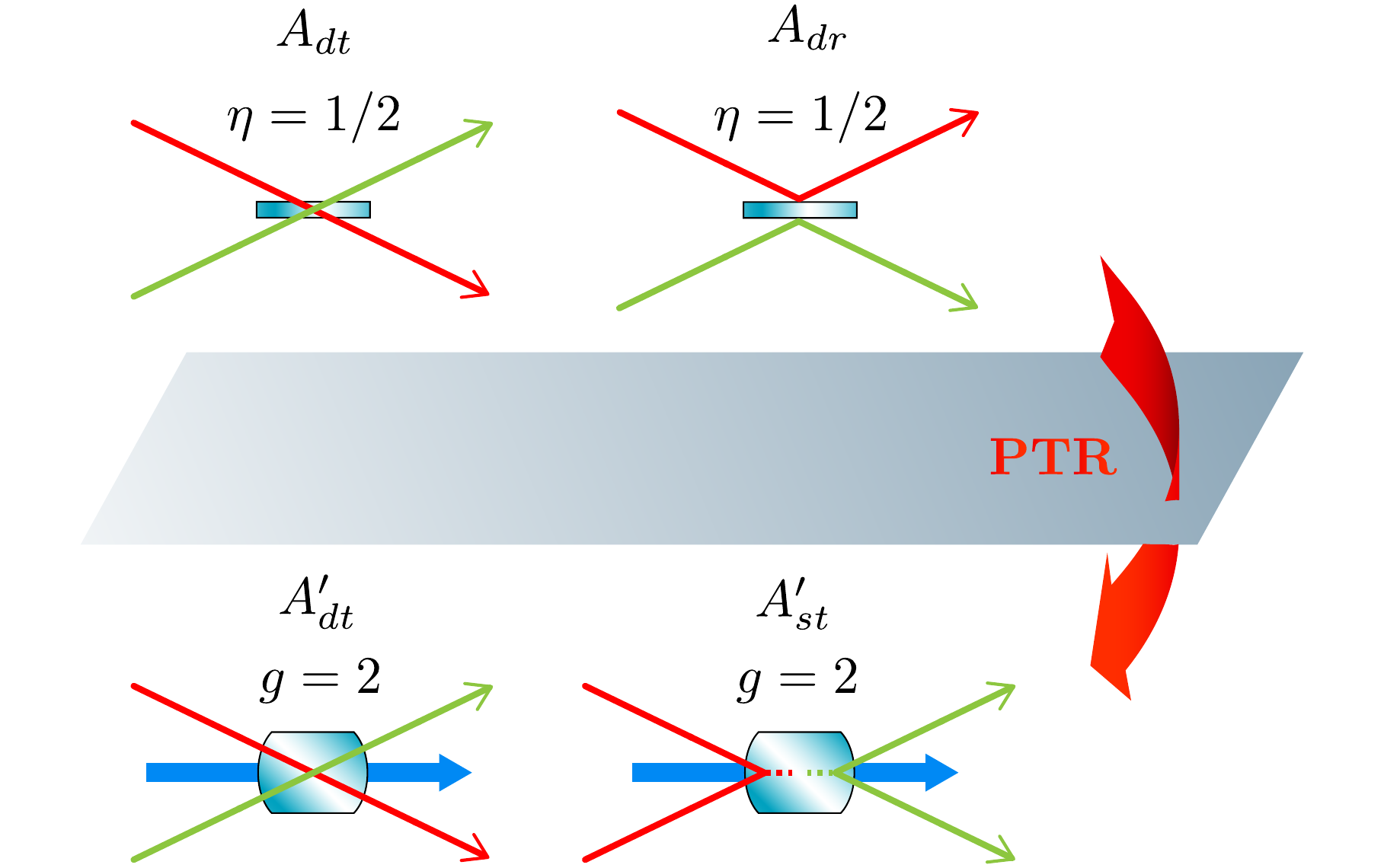}
\caption{The HOM effect (upper panel) is due to spacelike indistinguishability: the double-transmission path (of amplitude $A_\mathrm{dt}$) where the two photons are transmitted interferes destructively with the double-reflection path (of amplitude $A_\mathrm{dr}$) where the two photons are reflected. Exchanging the two photons in space leads to the HOM effect in a BS of transmittance $\eta=1/2$. According to PTR duality (by reversing the arrow of time in the second mode, see lower panel), the two interfering paths in a PDC correspond to the double-transmission event associated with amplitude  $A'_\mathrm{dt}$ (the two photons simply cross the PDC) and double-stimulated event of amplitude $A'_\mathrm{st}$ (the input photon pair is up-converted into the pump beam and another pump photon is down-converted into a new photon pair). Exchanging the photon pairs in time induces a timelike interferometric suppression in a PDC of gain $g=2$.
}
\label{fig-timelike-indistinguishability}
\end{figure}

The origin of the two-boson quantum interference effect that we predict can be traced back to boson indistinguishability, similarly as for the HOM effect albeit in a timelike version (involving bosons from the past and future). We first recall that the HOM effect originates from what can be viewed as ``spacelike'' indistinguishability, see Fig.~\ref{fig-timelike-indistinguishability} (upper panel). When two photons impinge on a BS of transmittance $\eta$, each photon has a probability amplitude $\sqrt{\eta}$ of being transmitted, so the double-transmission amplitude is $A_\mathrm{dt}= \eta$. In contrast, the probability amplitude of reflection is  $\sqrt{1-\eta}$ but with opposite signs for the two photons, so the double-reflection amplitude is  $A_\mathrm{dr}=-(1-\eta)$. Since a double-transmission event is indistinguishable from a double-reflection event, we must add probability amplitudes, leading to 
$ |A_\mathrm{dt} + A_\mathrm{dr}|^2 = P_\mathrm{coinc} (\eta)$. The double-transmission and double-reflection amplitudes exactly cancel out for $\eta=1/2$, which originates from the fact that an exchange of the two indistinguishable photons in space (which turns a double-transmission into a double-reflection event) cannot lead to any observable consequence.
\par

We now argue that it is the exchange of indistinguishable photons in time that is responsible for the interference effect in an amplifier, see Fig.~\ref{fig-timelike-indistinguishability} (lower panel).  When two photons impinge on a PDC with gain $g$, they can be both transmitted without triggering a stimulated event, which is dual to the double transmission in a BS (where $\eta$ is substituted by $1/g$). Hence, the double-transmission amplitude in a PDC is $A'_\mathrm{dt}= g^{-1/2} A_\mathrm{dt} = 1/g^{3/2}$. Another possible path giving rise to the coincident detection of two single photons is the combination of the stimulated annihilation and emission of a pair of photons, which is dual to the double reflection in a BS. This double-stimulated event admits a probability amplitude $A'_\mathrm{st} = g^{-1/2} A_\mathrm{dr} = - (g-1)/g^{3/2}$, where the minus sign results from the fact that the probability amplitude that the input pair disappears by stimulated annihilation and the probability amplitude that a new pair is created by stimulated emission have opposite signs. Again, since the double-transmission and double-stimulated events are indistinguishable, we must add their probability amplitudes and get $|A'_\mathrm{dt} + A'_\mathrm{st}|^2 = P'_\mathrm{coinc} (g)$, which vanishes when $g=2$. Roughly speaking, we cannot know whether the ``old'' photons have been replaced by ``new'' photons or have been left unchanged, which we dub timelike indistinguishability.
\par

\section*{Discussion and conclusion}

The role of time reversal in quantum physics has long been a fascinating subject of questioning (see, e.g.,  \cite{OreshkovCerf} and references therein), but the key idea of the present work is to consider a bipartite quantum system (two bosonic modes) with counter-propagating times. Incidentally, we note that the notion of  time reversal has been exploited in the context of defining separability criteria for bosonic systems \cite{Sanpera,Simon}, but this seems to be unrelated to PTR duality. Further, the link between time reversal and optical phase conjugation has been mentioned in the quantum optics literature, see e.g. \cite{Park}, but it exploits the fact that the complex conjugate of an electromagnetic wave is the time-reversed solution of the wave equation (the phase-conjugation time-reversal mirror concerns one mode only). The PTR duality introduced here bears some resemblance with an early model of lasers \cite{Glauber} based on the coupling of an ``inverted'' harmonic oscillator (having a negative frequency $\omega$) with a heat bath. The inverted harmonic oscillator ($\mathrm{e}^{-i\omega t} \to \mathrm{e}^{i\omega t}$) can indeed be viewed as a time-reversed harmonic oscillator. Quantum amplification in this model occurs from a PDC-like coupling of this inverted harmonic oscillator, whereas quantum damping follows from the BS-like coupling of a usual harmonic oscillator with the bath. The PTR duality is also reminiscent of Klyshko's so-called ``advanced-wave picture'' in PDC \cite{Klyshko}, which provides an interpretation of coincidence-based two-photon experiments: the wave that is detected by one of the detectors behind PDC can be viewed as resulting from an ``advanced wave'' emitted by the second detector, so that PDC acts as a mirror \cite{Arruda}. This picture may indeed be interpreted as a special case of Eq. (\ref{eq-fundamental-equation-2}), namely $\bra{0,0} \tms \ket{\psi,\phi} = g^{-1/2}  \bra{0,\phi^*} U^{\mathrm{BS}}_{1/g} \ket{\psi,0}$. In the limit where the gain $g\to \infty$, the two outputs of PDC can be viewed as the input $\psi$ and output $\phi^*$ of a fully reflecting (phase-conjugating) mirror with $\eta\to 0$.
\par

In this work, we have promoted partial time reversal as the proper way to approach the duality between passive and active bosonic transformations. 
As a compelling application of PTR duality, we have unveiled a hitherto unknown quantum interference effect, which is a manifestation of quantum indistinguishability for identical bosons in active transformations (spacelike indistinguishability, which is at the root of the HOM effect, transforms under PTR into timelike indistinguishability). The interferometric suppression of the coincident $\ket{1,1}$ term is induced by the indistinguishability between a photon pair originating from the past and a photon pair going to the future. Stated more dramatically, while the two photons may cross the amplification medium and be detected, the sole fact that they could instead be annihilated and replaced by two other photons makes the detection probability drop to zero when $g=2$.
\par

The experimental verification of this effect can be envisioned with present technologies (Materials and Methods G). A coincidence probability lower than 25\% would be sufficient to rule out a classical interpretation, which could in principle be reached with a moderate gain of 1.28 (Materials and Methods H).
Observing timelike two-photon interference in experiments involving active optical components would then be a highly valuable metrology tool given that the HOM dip is  commonly used today as a method to benchmark the reliability of single-particle sources and mode matching. 
More generally, the interference of many photons scattered over many modes in a linear optical network has generated a tremendous interest in the recent years, given the connection with the ``boson sampling'' problem, i.e., the hardness of computing the permanent of a random matrix \cite{Aaronson}, and technological progress in integrated optics now makes it possible to access large optical circuits (see, e.g., \cite{Walmsley}). In this context, it would be exciting to uncover new consequences of PTR duality and timelike interference.
\par

Finally, we emphasize that our analysis encompasses all bosonic Bogoliubov transformations, which are widespread in physics, appearing in quantum optics, quantum field theory, or solid-state physics, but also in black-hole physics or even in the Unruh effect (describing an accelerating reference frame). This suggests that timelike quantum interference may occur in various physical situations where identical bosons participate in such a transformation. Beyond bosons, let us point out an intriguing connection with the notion of ``crossing'' in quantum electrodynamics \cite{Gell-Mann,Crossing}. Crossing symmetry refers to a substitution rule connecting two scattering matrix elements that are related by a Wick rotation (antiparticles being turned into particles going backwards in time). For example, the scattering of a photon by an electron (Compton scattering) and the creation of an electron-positron pair by two photons are processes that are related to each other by such a substitution rule (see, e.g., \cite{QFT}). This is  in many senses analogous to the PTR duality described here: since a photon (or truly neutral boson) is its own antiparticle, we may view PTR duality as a substitution rule connecting the BS diagram to the PDC diagram.  
We hope that this connection with quantum electrodynamics may open up even broader perspectives.


\section*{Materials and Methods}

\subsection*{(A) Gaussian unitaries for a BS and PDC}
Passive and active Gaussian unitaries are effected by linear optical interferometry or parametric amplification, respectively \cite{RMP}.  The fundamental passive two-mode Gaussian unitary, namely the beam-splitter (BS) unitary
\begin{equation}
\bs = \exp \left[ \theta ( \ad \b - \a \bd ) \right], \quad \eta = \cos^2 \theta ,
\label{eq-BS-unitary}
\end{equation}
effects an energy-conserving linear coupling between modes $\a$ and $\b$, and acts in the Heisenberg picture as 
\begin{equation}
\begin{aligned}
\a \to \a' & =   \bsd \, \hat a \, \bs =  \a \, \cos\theta  + \b \, \sin\theta , \\
\b \to \b' & =   \bsd \, \hat b \, \bs = - \a \, \sin\theta  + \b \, \cos\theta ,
\end{aligned}
\label{eq-BS-Heisenberg}
\end{equation}
where $\hat a$ and $\hat b$ are the mode operators and $\eta$ is the transmittance ($0\le \eta \le 1$). 
Similarly, the unitary 
\begin{equation}
\tms = \exp \left[ r ( \ad \bd - \a \b ) \right], \quad g = \cosh^2 r,
\label{eq-TMS-unitary}
\end{equation}
models the generation of pairs of entangled photons by parametric down-conversion (PDC) due to the optical pumping of a nonlinear crystal. It transforms the mode operators according to the Bogoliubov transformation
\begin{equation}
\begin{aligned}
\a \to \a' & = \tmsd \, \hat a \, \tms  =  \a \, \cosh r  +  \bd \, \sinh r    , \\
\b \to \b' & = \tmsd \, \hat b \, \tms  =  \ad \, \sinh r + \b \, \cosh r   ,
\end{aligned}
\label{eq-TMS-Heisenberg}
\end{equation}
where $g$ is the parametric gain ($g\ge 1$) and $r$ is the squeezing parameter. The photon-number difference is conserved by the PDC transformation, namely $\a'^\dagger \a' - \b'^\dagger \b' = \ad\a-\bd\b$.
\par

The action of $\bs$ on Fock states can be expressed by using the decomposition of the exponential
\begin{eqnarray}
\bs &=  \exp \left( \ad \b \tan \theta \right)   \,  \left( \frac{1}{\cos \theta}\right)^{\ad\a-\bd\b}   \nonumber \\
&\times  \exp \left( - \a \bd \tan \theta \right)  \,  .
\label{eq-disentanglement-BS}
\end{eqnarray}
Alternatively, it can easily be computed by using the canonical transformation, Eq.~(\ref{eq-BS-Heisenberg}). For example, when $n$ photons impinge on one of the input ports, each photon may be transmitted or reflected, so we get the binomial state
\begin{eqnarray}
\!\!\!\!\!\!  \bs \ket{n,0} \!\! &=& \!\!  \frac{ (\ad \, \cos\theta  - \bd \, \sin\theta)^n }{ \sqrt{n!} } \, \bs \ket{0,0}   \nonumber \\
 &=& \!\! \sum_{k=0}^n \binom{n}{k}^{1/2}    (\cos\theta)^k (-\sin\theta)^{n-k} \, \ket{k,n-k} 
\end{eqnarray}
or 
\begin{eqnarray}
\!\!\!\!\!\!  \bs \ket{0,n} \!\! &=& \!\!  \frac{ (\ad \, \sin\theta  + \bd \, \cos\theta)^n }{ \sqrt{n!} } \, \bs \ket{0,0}   \nonumber \\
 &=& \sum_{k=0}^n \binom{n}{k}^{1/2}    (\sin\theta)^k (\cos\theta)^{n-k} \, \ket{k,n-k}   .
\end{eqnarray}
The action of $\tms$ on Fock states can be conveniently expressed by use of the disentanglement formula,
\begin{eqnarray}
\tms &=  \exp \left( \ad \bd \tanh r \right)   \,  \left( \frac{1 }{ \cosh r } \right)^{1+\ad\a+\bd\b}   \nonumber \\
&\times  \exp \left( - \a \b \tanh r \right)    .
\label{eq-disentanglement-PDC}
\end{eqnarray}
For example, the two-mode squeezed vacuum state is obtained by applying $\tms$ onto the vacuum state, namely
\begin{equation}
\tms | 0,0 \rangle = \frac{1}{\cosh r} \sum_{n=0}^{\infty} \tanh^n r  \, | n,n \rangle ,
\label{eq-TMVS}
\end{equation}
where the $n$th term in the right-hand side corresponds to the stimulated emission of $n$ photon pairs.
In the more general case where $m$ photons are impinging into mode $\a$ (with vacuum in the other input mode), we have
\begin{eqnarray}
 \tms \ket{m,0} &=&  \frac{1 }{ (\cosh r)^{m+1} } \sum_{n=0}^\infty \binom{n+m}{m}^{1/2}   \nonumber \\
 && ~~~~~~ \times \tanh^n r \, \ket{n+m,n}  ,
\label{eq-PDC-m-0-photons}
\end{eqnarray}
where each term corresponds again to the stimulated emission of $n$ photon pairs, with $m$ extra photons traveling in mode $\a$.
\par

\begin{figure}[t]
\centering
\includegraphics[trim = 0cm 0cm 0cm 0cm, clip, width=0.8\linewidth, page=6]{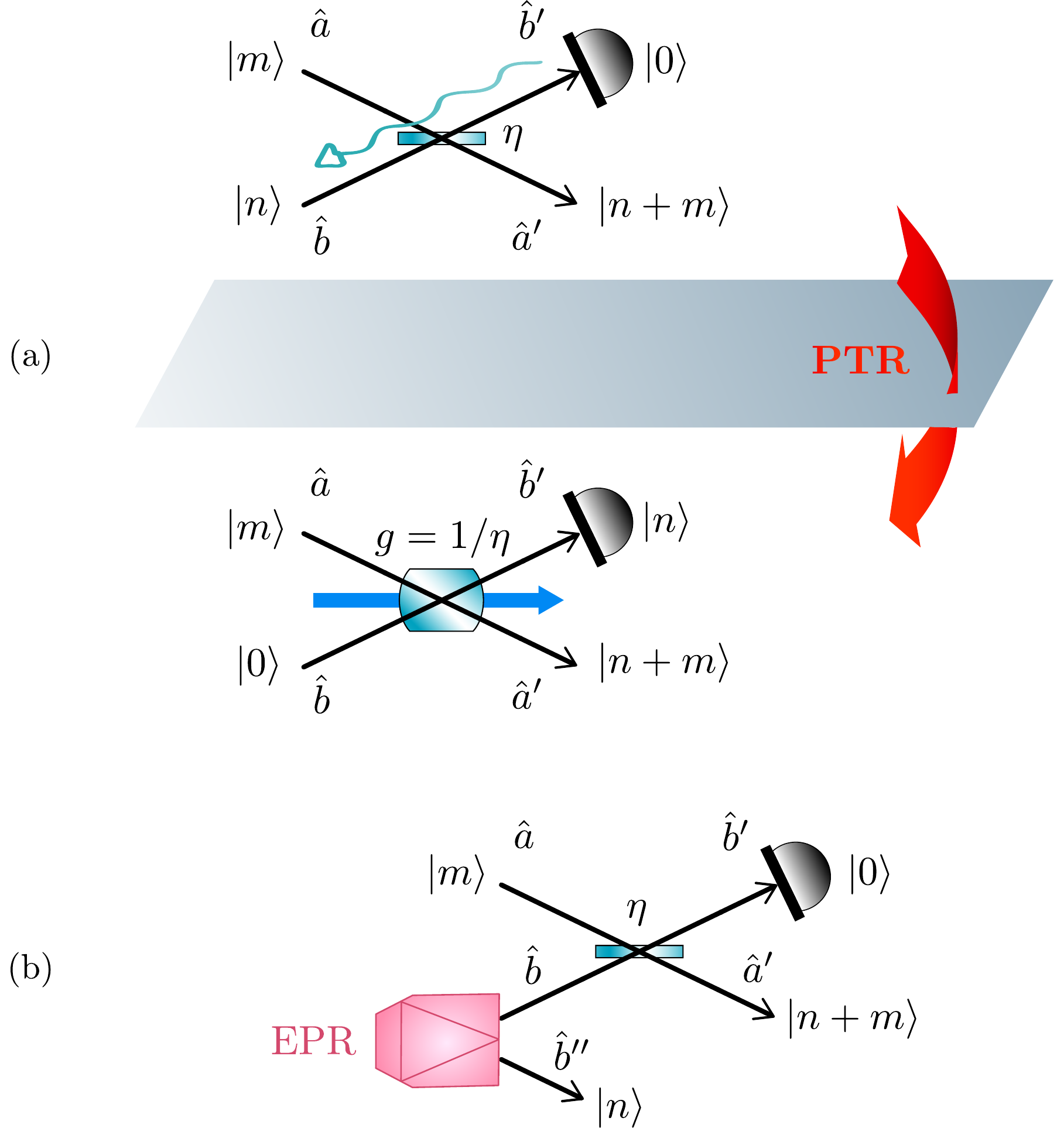}
\caption{(a) PTR duality in a more general case with $m$ and $n$ photons impinging on modes $\a$ and $\b$ of a BS. By conditioning the output mode $\b'$  on vacuum and reversing the arrow of time, we get the same transition probability amplitude (up to a constant) as for a PDC of gain $g=1/\eta$ with input state $\ket{m,0}$ and output state $\ket{n+m,n}$. (b)~Corresponding operational scheme using an entangled (EPR) state at the input of mode $\b$. This makes it possible to access the output retrodicted state on mode $\b''$.
}
\label{fig-BS-n-and-m-photons}
\end{figure}

\subsection*{(B) Example of PTR}
We illustrate the PTR duality between a BS and PBC by considering the additional example of a BS with $m$ photons impinging on input port $\a$ and $n$ photons impinging on input port $\b$, see Fig.~\ref{fig-BS-n-and-m-photons}a. If we condition on the vacuum on mode $\b'$ at the output of the BS, the (unnormalized) conditional output state of mode $\a'$ is 
\begin{equation}
\ket{\phi_n} = \binom{n+m}{m}^{1/2} \sin^n \!\theta  \, \cos^m \!\theta \, \ket{n+m} ,
\end{equation}
hence we have the probability amplitude 
\begin{equation}
\bra{n+m,0} \bs \ket{m,n} =  \binom{n+m}{m}^{1/2} \sin^n \!\theta  \, \cos^m \!\theta   \,  .
\label{eq-amplitude-BS-n+m}
\end{equation}
The special case $m=0$ is considered in the main text (Fig.~\ref{fig-BS-under-PTR}) and corresponds to $\ket{\phi_n} =  \sin^n \! \theta \, \ket{n}$.
By reversing the arrow of time on mode $\b$, we compare the probability amplitude of Eq. (\ref{eq-amplitude-BS-n+m})
to the probability amplitude 
\begin{equation}
\bra{n+m,n} \tms \ket{m,0} =  \binom{n+m}{m}^{1/2} \frac {\tanh^n r} {(\cosh r)^{m+1}}
\label{eq-amplitude-PDC-n+m}
\end{equation}
for a PDC of gain $g$, see Eq.~(\ref{eq-PDC-m-0-photons}). Now if we make the substitution $\sin \theta = \tanh r$ and $\cos \theta = (\cosh r)^{-1}$, which is equivalent to $g=1/\eta$, we confirm that Eqs. (\ref{eq-amplitude-BS-n+m}) and (\ref{eq-amplitude-PDC-n+m}) are dual under PTR, namely
\begin{equation}
\bra{n+m,n} \tms \ket{m,0} = \frac{1 }{ \sqrt {g}} \, \bra{n+m,0} U^{\mathrm{BS}}_{1/g}  \ket{m,n} .
\end{equation}
As shown in Fig.~\ref{fig-BS-n-and-m-photons}b, if the input mode $\b$ is entangled (that is, if we use the entangled state $\ket{\Psi}_{b,b''}\propto \sum_{n=0}^\infty \ket{n,n}$), we get the output state $\ket{\Psi}_{a',b''} \propto \sum_{n=0}^\infty \ket{\phi_n,n}$, which is precisely proportional to the output of a PDC when the signal and idler modes are initially in states $\ket{m}$ and $\ket{0}$, respectively, see Eq.~(\ref{eq-PDC-m-0-photons}).
\par

\begin{table}[t]
\centering
\begin{tabular}{l l }
\hline
BS & \quad  PDC \\
\hline
$\bra{0,0}  U^{\mathrm{BS}}_{\eta} \ket{0,0} = 1$ &\quad $\bra{0,0}  \tms  \ket{0,0}  =  1/ \sqrt {g} $ \\  
$\bra{1,0}  U^{\mathrm{BS}}_{\eta} \ket{1,0} = \sqrt{\eta}$  &\quad    $\bra{1,0} \tms \ket{1,0} = 1/ g$ \\
$\bra{0,1}  U^{\mathrm{BS}}_{\eta} \ket{0,1} = \sqrt{\eta}$   &\quad   $\bra{0,1} \tms \ket{0,1} = 1/ g$  \\
$\bra{0,1}  U^{\mathrm{BS}}_{\eta} \ket{1,0} = -\sqrt {1-\eta}$   &\quad   $\bra{0,0} \tms \ket{1,1} = - \sqrt {g-1} / g $ \\
$\bra{1,0}  U^{\mathrm{BS}}_{\eta} \ket{0,1} = \sqrt {1-\eta}$    &\quad     $\bra{1,1} \tms \ket{0,0} =  \sqrt {g-1} / g  $ \\
\hline
\end{tabular}
\caption{Illustration of PTR duality with few photons. The second column (PDC) is obtained from the first column (BS) by time-reversing mode $\b$, substituting $\eta$ with $1/g$, and dividing by the factor $\sqrt{g}$. The first row explains the latter factor: vacuum is obviously conserved in a BS, while PDC implies the stimulated emission of photon pairs (hence, the probability of keeping vacuum is strictly lower than 1). The second and third rows correspond to the transmission of a single photon through the BS or PDC. The fourth and fifth rows correspond to the reflection of a single photon by the BS or the stimulated annihilation (fourth row) or emission (fifth row) of a photon pair by PDC.}
\label{table-PTR-duality}
\end{table}

\subsection*{(C) Proof of PTR duality}
\begin{figure}[t]
\centering
\includegraphics[trim = 0cm 0cm 0cm 0cm, clip, width=0.8 \linewidth, page=7]{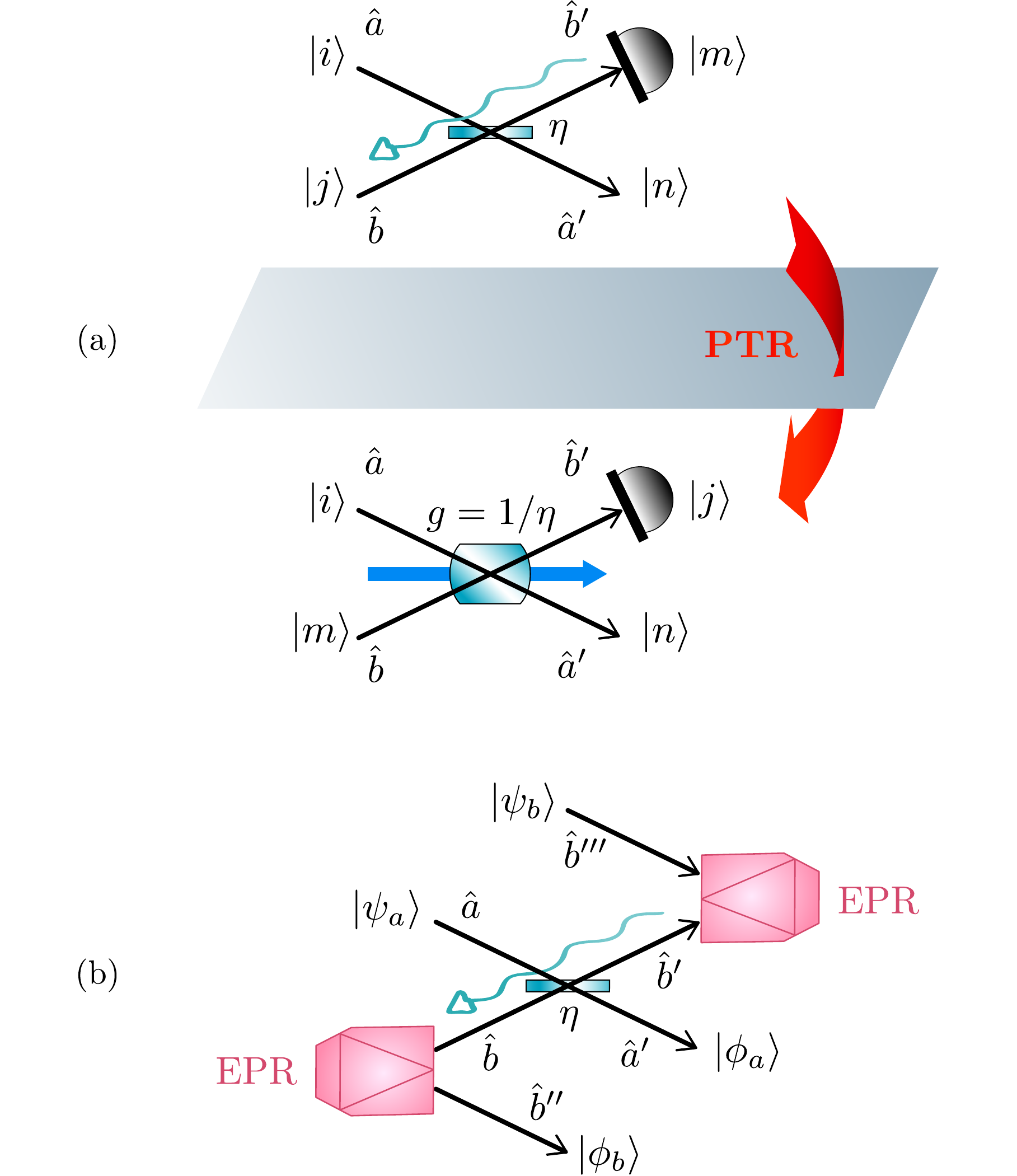}
\caption{(a) General statement of the PTR duality as expressed in Eq.~(\ref{eq-fundamental-equation}), when $i$ and $j$ photons impinge on input modes $\a$ and $\b$ of a BS, while $n$ and $m$ photons are detected in output modes $\a'$ and $\b'$. We recover a PDC with input $\ket{i,m}$  and output $\ket{n,j}$.
(b)  Corresponding operational scheme, where a PDC is emulated with a BS. The input state $\ket{\psi_a}$ of mode $\hat a$ simply propagates through the BS and is projected onto $\ket{\phi_a}$. The input state $\ket{\psi_b}$ of mode $\b'''$ is converted, by projection on an EPR state, into the retrodicted state $\ket{\psi_b^*}$ of mode $\b'$, which backpropagates through the BS and is projected onto $\ket{\phi_b^*}$ in mode $\b$. We obtain the corresponding output state $\ket{\phi_b}$ on mode $\b''$ by using an EPR pair. We thus recover a PDC with input $\ket{\psi_a,\psi_b}$  and output $\ket{\phi_a,\phi_b}$, as encapsulated by Eq.~(\ref{eq-fundamental-equation-2}).
}
\label{fig-general-PTR-duality}
\end{figure}

The PTR duality is illustrated in Table~\ref{table-PTR-duality} for few photons. As expressed by Eq.~(\ref{eq-fundamental-equation}), it can be viewed as a consequence of partial transposition of the state of mode $\b$ (leaving mode $\a$ unchanged), namely the fundamental relation
\begin{equation}
\left(  \tms  \right)^{T_b}  =  \frac{1 }{ \sqrt {g}} \,  U^{\mathrm{BS}}_{1/g}  
\label{eq-fundamental-partial transpose}
\end{equation}
where $T_b$ stands for transposition in the Fock basis of the Hilbert space associated with mode $\b$. This is sketched in Fig.~\ref{fig-general-PTR-duality}a and can also be interpreted by comparing the unitaries $\bs$ and $\tms$ in Eqs.~(\ref{eq-BS-unitary}) and (\ref{eq-TMS-unitary}), or their corresponding decompositions, Eqs.~(\ref{eq-disentanglement-BS}) and (\ref{eq-disentanglement-PDC}).
In general terms, we may say that the (passive) linear coupling of two bosonic modes is dual, under PTR, to an (active) Bogoliubov transformation, which is expressed as
\begin{equation}
\bra{\phi_a,\phi_b}  \tms  \ket{\psi_a,\psi_b}  =  \frac{1 }{ \sqrt {g}} \, \bra{\phi_a,\psi_b^*}  U^{\mathrm{BS}}_{1/g} \ket{\psi_a,\phi_b^*} 
\label{eq-fundamental-equation-2}
\end{equation}
for any states $\psi_a$, $\psi_b$, $\phi_a$, and $\phi_b$, where $*$ denotes the complex conjugation in the Fock basis. In Fig. \ref{fig-general-PTR-duality}b, the corresponding operational scheme is depicted, relying on the preparation of an entangled (EPR) state at the input of mode $\b$ and the projection onto an entangled (EPR) state at the output of mode $\b'$.
\par

We now prove PTR duality by reexpressing Eq. (\ref{eq-fundamental-partial transpose}) in the Heisenberg picture, namely
\begin{equation}
\begin{aligned}
V^{T_a} \, \a \, V^{T_b} &=  (\a \, \cos\theta  + \b \, \sin\theta) /g   \,  , \\
V^{T_a} \, \b \, V^{T_b} &=  (\b \, \cos\theta - \a \, \sin\theta) /g     \,  ,
\end{aligned}
\label{eq-transpose}
\end{equation}
where $V\equiv \tms$. Note that $\hat a$ and $\b$ are real operators in Fock basis, hence the same is true for $V$. Therefore, we have $\ad=a^T$, $\bd=b^T$, and  $V^\dagger = V^T$ where $T$ stands for transposition in Fock basis. Obviously, we have $(V^{T_b})^T = V^{T_a}$.
By using the identity $((A\otimes B) M)^{T_b} = (A\otimes \mathbbm{1}) M^{T_b} (\mathbbm{1} \otimes B^T)$, where $\mathbbm{1}$ represents the identity operator, 
we express $\a_\mathrm{PTR} \equiv V^{T_a} \, \a \, V^{T_b} $ as
\begin{eqnarray}
\lefteqn{  V^{T_a} \, (\a \, V)^{T_b}  = V^{T_a} \, (V V^\dagger \a \, V)^{T_b}  }  \hspace{1cm}  \nonumber \\
 &=& V^{T_a} \, [V (\a \, \cosh r + \bd \sinh r)]^{T_b} \nonumber \\
 &=& V^{T_a} V^{T_b} \, \a \, \cosh r + V^{T_a} \b \, V^{T_b} \, \sinh r   \, .
 \label{eq-VTa-a-VTb-1} 
\end{eqnarray}
Equivalently, using $((A\otimes B) M)^{T_a} = (\mathbbm{1} \otimes B) M^{T_a} (A^T \otimes \mathbbm{1})$, 
we may also reexpress $\a_\mathrm{PTR}$ as 
\begin{eqnarray}
\lefteqn{  (\ad V)^{T_a} \, V^{T_b}  =  (V V^\dagger \ad V)^{T_a} \, V^{T_b}  }  \hspace{1cm} \nonumber \\
 &=&  [V (\ad \cosh r + \b \, \sinh r)]^{T_a} \, V^{T_b}  \nonumber \\
 &=& \a \,  V^{T_a} V^{T_b}  \cosh r + V^{T_a} \b \, V^{T_b} \, \sinh r  \, .
\label{eq-VTa-a-VTb-2}
\end{eqnarray}
Defining the operators $\b_\mathrm{PTR} \equiv V^{T_a} \, \b \, V^{T_b}$, $W\equiv V^{T_a} V^{T_b}$, and identifying Eq. (\ref{eq-VTa-a-VTb-1}) with Eq. (\ref{eq-VTa-a-VTb-2}), we see that 
\begin{eqnarray}
\a_\mathrm{PTR} = \a \, W  \cosh r + \b_\mathrm{PTR} \, \sinh r,  \quad [W,\a]=0  \, .
\label{eq-a_PTR}
\end{eqnarray}
We can perform a similar development starting from $\b_\mathrm{PTR}$, resulting in
\begin{eqnarray}
\b_\mathrm{PTR}  = \b \, W \cosh r -  \a_\mathrm{PTR} \, \sinh r,  \quad [W,\b]=0   \, .
\label{eq-b_PTR}
\end{eqnarray}
Now, solving Eqs. (\ref{eq-a_PTR}) and (\ref{eq-b_PTR}) for $\a_\mathrm{PTR}$ and $\b_\mathrm{PTR}$, we get
\begin{equation}
\begin{aligned}
\a_\mathrm{PTR} =  \a \, W \cos \theta + \b \, W \sin \theta    \,  ,  \\
\b_\mathrm{PTR}  = \b \, W \cos \theta -  \a \, W \sin \theta    \,  ,
\end{aligned}
\label{eq-transpose-bis}
\end{equation}
where we have made the substitutions $\cosh r=(\cos \theta)^{-1}$ and $\sinh r=\tan \theta$.
\par

Similar equations can be derived starting from $V^{T_a} \, \ad \, V^{T_b}$ and $V^{T_a} \, \bd \, V^{T_b}$, which imply that the operator $W$ also commutes with $\ad$ and $\bd$. Hence, $W$ is a scalar (proportional to~$\mathbbm{1}$) and it is sufficient to compute its diagonal matrix element in an arbitrary state, e.g.,  $\ket{0,0}$. Recalling that $V^\dagger = V^T$, we have $W= (V^\dagger)^{T_b} V^{T_b}$, so that 
\begin{eqnarray}
\bra{0,0}W\ket{0,0} &=& \sum_{k,l=0}^\infty  \bra{0,0} (V^\dagger)^{T_b} \ket{k,l} \, \bra{k,l} V^{T_b} \ket{0,0}  \nonumber \\
&=& \sum_{k,l=0}^\infty \bra{0,l} V^\dagger \ket{k,0} \, \bra{k,0} V \ket{0,l} \nonumber \\
&=& |\bra{0,0} V \ket{0,0}|^2 \nonumber \\
&=& 1/g \, .
\end{eqnarray}
Substituting $W$ with $1/g$ in Eqs. (\ref{eq-transpose-bis}), we get Eqs. (\ref{eq-transpose}), which concludes the proof of Eq. (\ref{eq-fundamental-partial transpose}).
\par

Note that PTR duality can be reexpressed by using the identity
\begin{equation}
    \begin{aligned}
        & \mathrm{Tr} \left[ U_{ab} \left( \hat{X}_a \otimes \hat{X}_b \right) U_{ab}^{\dagger} \left( \hat{Y}_a \otimes \hat{Y}_b \right) \right] \\
        & \vspace{3cm} = \mathrm{Tr} \left[ U_{ab}^{T_b} \left( \hat{X}_a \otimes  \hat{Y}_b^T \right) \left(U_{ab}^{T_b}\right)^{\dagger} \left( \hat{Y}_a \otimes  \hat{X}_b^T \right) \right],
    \end{aligned}
    \label{eq:partialTransposeGeneral}
\end{equation}
which is valid for any joint unitary $U_{ab}$ and for any operators $\hat{X}_{a(b)}$ and $\hat{Y}_{a(b)}$ acting on mode $\a$ ($\b$). Plugging $U_{ab}=\tms$ into Eq. (\ref{eq:partialTransposeGeneral}) and using Eq. (\ref{eq-fundamental-partial transpose}) implies the general relation
\begin{equation}
    \begin{aligned}
        & \mathrm{Tr} \left[ \tms \left( \hat{X}_a \otimes \hat{X}_b \right) \tmsd \left( \hat{Y}_a \otimes \hat{Y}_b \right) \right] \\
        & \vspace{3cm} = \frac{1}{g} \, \mathrm{Tr} \left[ U^{\mathrm{BS}}_{1/g}  \left( \hat{X}_a \otimes \hat{Y}_b^{T} \right) U^{\mathrm{BS} \dagger}_{1/g}   \left( \hat{Y}_a \otimes \hat{X}_b^{T} \right) \right]  .
    \end{aligned}
    \label{eq:partialTransposeGeneralQI}
\end{equation}
This equation is needed to interpret PTR in the context of the retrodictive picture of quantum mechanics. 
\par

\begin{figure}[t]
\centering
    \includegraphics[trim = 0cm 0cm 0cm 0cm, clip, width=0.8 \linewidth]{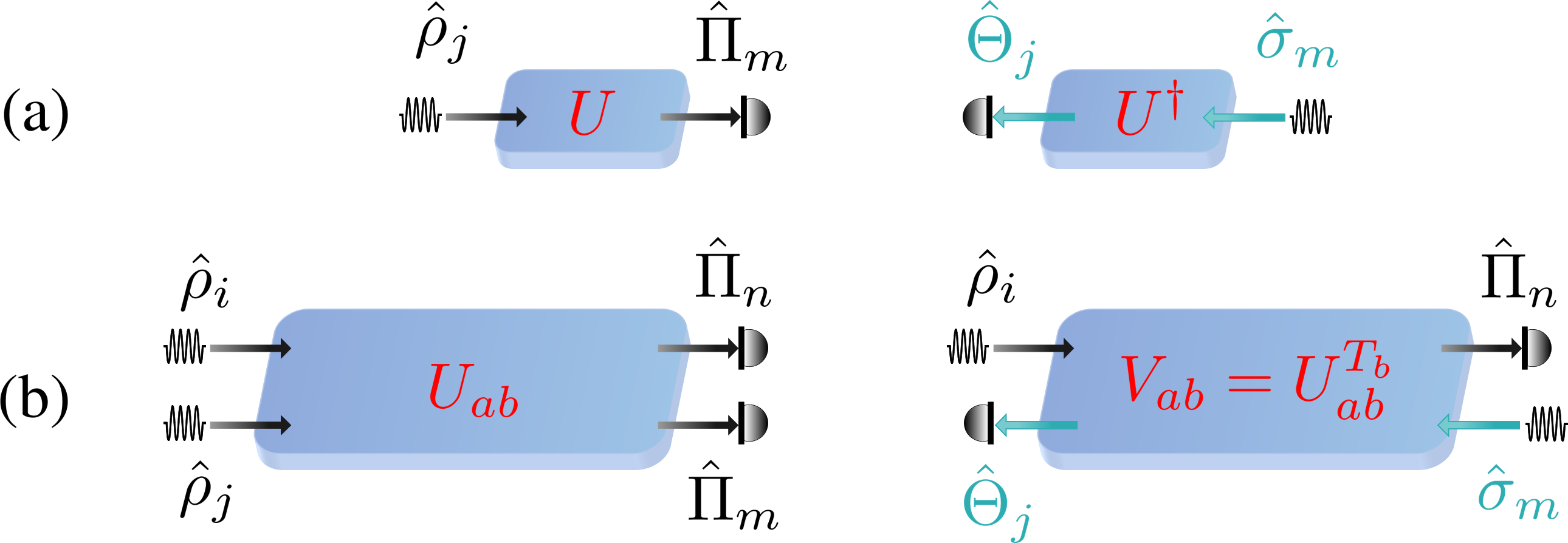}
    \caption{(a) Predictive (left) and retrodictive (right) pictures, describing the same experiment where state $\hat \rho_j$ is prepared, evolves according to $U$, and leads to measurement outcome associated with $\hat \Pi_m$. The probability $\mathbb{P}(j|m)$ can be written as resulting from the retrodictive state $\hat\sigma_m$ back propagating according to $U^{\dagger}$, followed by measurement $\hat\Theta_j$. (b) Predictive (left) and intermediate (right) pictures for a bipartite system. In the latter picture, associated with PTR, the retrodictive state of subsystem $b$ propagates backwards in time, while the predictive state of subsystem $a$ propagates forwards in time. If $U_{ab}$ is a unitary, $V_{ab} = U_{ab}^{T_b}$ is not necessarily proportional to a unitary;  however, it is the case when considering the BS vs PDC duality.    }
    \label{fig:retroPic}
\end{figure}

\subsection*{(D) Retrodictive picture of quantum mechanics}
In the usual, predictive approach of quantum mechanics, one deals with the preparation of a quantum system followed by its time evolution and ultimately its measurement. Specifically, one uses the prior knowledge on the state $\hat \rho_j$ (prepared with probability $p_j$) in order to make predictions about the outcomes of a measurement $\{\hat \Pi_m\}$. If the state evolves according to unitary $U$ before being measured, Born's rule provides the conditional probabilities $\mathbb{P}(m|j) = \mathrm{Tr}[U \hat\rho_j U^{\dagger} \hat\Pi_m]$. In contrast, in the retrodictive approach of quantum mechanics \cite{Aharonov}, one postselects the instances where a particular measurement outcome $m$ was observed and one focuses on the probability of the preparation variable $j$ conditionally on this measurement outcome. This can be interpreted as if the actually measured state had propagated backwards in time to the preparer (see Fig.~\ref{fig:retroPic}a). Specifically, one associates a ``retrodicted'' state $\hat\sigma_m$ to the observed outcome $m$ and makes retrodictions about the preparation by evolving $\hat\sigma_m$ according to $U^\dagger$ and applying a measurement $\{ \hat\Theta_j \}$, whose outcome $j$ discriminates the prepared state $\hat\rho_j$. In the simplest case (to which we restrict here) where there is no \textit{a priori} information about the source, one sets $\sum_j p_j\, \hat\rho_j \propto \mathbbm{1}$. Then, by defining
\begin{equation}
\hat\sigma_m = \frac{\hat\Pi_m }{ \mathrm{Tr}[\hat\Pi_m]} ,  \qquad
\hat\Theta_j = p_j\, \hat\rho_j,
\end{equation}
one recovers the expected properties of a state ($\hat\sigma_m \ge 0$, $\mathrm{Tr} \, \hat\sigma_m =1$) and of  a measurement ($\hat\Theta_j \ge 0$,  $\sum_j \hat\Theta_j =\mathbbm{1}$). Now, applying Born's rule to the retrodicted state $\hat\sigma_m$ having evolved according to $U^{\dagger}$ followed by a measurement $\hat\Theta_j$, we get
\begin{equation}
\mathrm{Tr}[U^{\dagger} \hat\sigma_m U \hat\Theta_j] = \frac{  p_j  \mathrm{Tr}[\hat\rho_j U^{\dagger} \hat\Pi_m U] }{  \sum_j  p_j  \mathrm{Tr}[U \hat\rho_j U^{\dagger} \hat\Pi_m] } = \mathbb{P}(j|m) ,
\end{equation}
which is consistent with Born's rule in the forward direction combined with Bayes' rule.

The retrodictive picture can be successfully exploited in different situations, for example to characterize the quantum properties of an optical measurement device \cite{Amri}, but it is always used \textit{in lieu of} the predictive picture. Here, we instead combine it with the predictive picture in order to properly define PTR duality and describe a composite system that is propagated partly forwards and partly backwards in time, as represented in Fig.~\ref{fig:retroPic}b (right). Specifically, we consider a composite system prepared in a product state $ \hat\rho^a_i \otimes \hat\rho^b_j$, then undergoing a unitary evolution $U_{ab}$ followed by a product measurement $\{ \hat\Pi^a_n \otimes \hat\Pi^b_m \}$. In the fully predictive picture, see Fig.~\ref{fig:retroPic}b (left), the conditional probabilities are given by
\begin{equation}
    \mathbb{P}(n,m|i,j) = \mathrm{Tr}\left[ U_{ab} \left( \hat\rho^a_i \otimes \hat\rho^b_j \right) U_{ab}^{\dagger} \left( \hat\Pi^a_n \otimes \hat\Pi^b_m \right) \right].
\end{equation}
In our intermediate picture, we postselect the instances where a particular measurement outcome $m$ was observed in subsystem $b$ when subsystem $a$ was prepared in state $\hat\rho^a_i$, and consider the joint probability of the preparation variable $j$ of subsystem $b$ together with the measurement outcome $n$ of subsystem $a$. Bayes' rule yields
\begin{eqnarray}
        \mathbb{P}(n,j|i,m) & = & \frac{\mathbb{P}(i) \, \mathbb{P}(j) \, \mathbb{P}(n,m|i,j)}{\mathbb{P}(i) \sum_{n'j'} \mathbb{P}(j') \, \mathbb{P}(n',m|i,j')} \nonumber \\
        & = & \frac{p^b_j \, \mathbb{P}(n,m|i,j)}{\sum_{n'j'} p^b_{j'} \, \mathbb{P}(n',m|i,j')}  ,
    \label{eq:BayesPartialRetro}
\end{eqnarray}
where $p^b_j$ is the probability that subsystem $b$ is prepared in state $\hat\rho^b_j$. Similarly as before, without information about the source, we set $\sum_j p^b_j\, \hat\rho^b_j \propto \mathbbm{1}$. We associate a retrodicted state $\hat{\sigma}^b_m$ to the observed outcome $m$ on subsystem $b$, while subsystem $a$ is still considered in the initial state $\hat{\rho}^a_i$. Similarly, we make retrodictions about the preparation of subsystem $b$ by applying measurement $\{ \hat{\Theta}^b_j \}$, whose outcome $j$ discriminates the prepared state $\hat{\rho}^b_j$, while still measuring subsystem $a$ according to $\{\hat\Pi^a_n\}$. Using identity (\ref{eq:partialTransposeGeneral}) and defining 
\begin{equation}
    \hat{\sigma}^b_m = \frac{\left( \hat{\Pi}^b_m \right)^T}{\mathrm{Tr}\left[ \hat{\Pi}^b_m  \right]} ,  \qquad
    \hat{\Theta}^b_j = p^b_j \, \left( \hat{\rho}^b_j \right)^T,
\end{equation}
which are easily seen to behave as a proper state and measurement,  we may reexpress  Eq. (\ref{eq:BayesPartialRetro}) as
\begin{equation}
    \begin{aligned}
        \mathbb{P}(n,j|i,m) & = \mathrm{Tr}\left[ V_{ab} \left( \hat\rho^a_i \otimes \hat{\sigma}^b_m \right) V_{ab}^{\dagger} \left( \hat\Pi^a_n \otimes \hat{\Theta}^b_j \right) \right],
    \end{aligned}
    \label{eq:BayesPartialRetro_2}
\end{equation}
where $V_{ab} \equiv  U_{ab}^{T_b}$. This can be viewed as the evolution of state $\hat{\rho}^a_i \otimes \hat{\sigma}^b_m$ according to $V_{ab}$, followed by the measurement of $\hat\Pi^a_n \otimes \hat{\Theta}^b_j$. In other words, in Eq. (\ref{eq:BayesPartialRetro_2}), the predictive picture is used for subsystem $a$, while the retrodictive picture is used for subsystem $b$.
\par

In our analysis of a BS under PTR, we have $U_{ab} = \tms$ and  $V_{ab} = U^{\mathrm{BS}}_{1/g} / \sqrt{g}$, so that Eq. (\ref{eq:partialTransposeGeneral}) reduces to Eq. (\ref{eq:partialTransposeGeneralQI}). Hence, $V_{ab}$ is unitary (up to a constant)  and can be interpreted as the propagation of the retrodicted state of mode $\hat b$ backwards in time through the BS, while the predictive state of mode $\hat a$ normally propagates forwards in time through the BS. According to Eq. (\ref{eq:BayesPartialRetro_2}), the joint state is then shown to evolve according to a PDC. Note that it is not always possible to construct an operator $V_{ab}$ that is proportional to a unitary operator, as it is the case here.
\par

\subsection*{(E) Two-photon interference in a BS and PDC}
The HOM effect can be simply understood by  calculating the probability amplitude for coincident detection
\begin{equation}
c \equiv \bra{1,1}  \bs \ket{ 1,1} =\bra{0,0}\a \b \, \bs  \, \ad \bd \ket{0,0}
\end{equation}
at the output of a BS. By using Eq. (\ref{eq-BS-Heisenberg}), it is simple to rewrite it as $c = \cos 2\theta$, which yields the coincidence probability
\begin{equation}
P_\mathrm{coinc} (\eta) = |c|^2 = (2\eta-1)^2 \, .
\label{eq-coinc-BS}
\end{equation}
If the transmittance $\eta=1/2$, no coincident detections can be observed as a result of destructive interference.
\par

Now, we examine the corresponding quantum interferometric suppression in a PDC and its dependence in the parametric gain $g$. Let us calculate the probability amplitude for coincident detection
\begin{equation}
c' \equiv \bra{1,1}  \tms \ket{ 1,1} = \bra{0,0}V V^\dagger \a V V^\dagger \b V \ad \bd \ket{0,0}  
\end{equation}
where $V$ is a short-hand notation for $\tms$. Thus, we get the scalar product between $V^\dagger \ket{0,0}$, which is the state of Eq.~(\ref{eq-TMVS}) up to changing the sign of $r$, and the ket  $V^\dagger \a V V^\dagger \b V \ad \bd \ket{0,0}$, which can be reexpressed as
\begin{equation}
\cosh^2 r \, \ket{0,0}+ 3 \cosh r \sinh r \, \ket{1,1} + 2 \sinh^2 r \, \ket{2,2}
\end{equation}
by use of Eq.~(\ref{eq-TMS-Heisenberg}). This gives $c' = (1- \sinh^2 r)/ \cosh^3 r$, so that the probability for coincidence is written as
\begin{equation}
P'_\mathrm{coinc} (g) = |c'|^2 = (2-g)^2 / g^3  .
\label{eq-coinc-PDC}
\end{equation}
If the gain $g=2$, the probability for coincident detections fully vanishes. 
More generally, the joint state at the output of a PDC of arbitrary gain $g$ when the input state is $\ket{1,1}$ reads
\begin{equation}
\tms \ket{ 1,1}=\sum_{n=0}^\infty \frac{  (\sinh r)^{n-1}  } {(\cosh r)^{n+2}}  \left( n - \sinh^2 r \right)  \ket{n,n}  \, .
\label{eq-output-coshr-sinhr}
\end{equation}
A parametric gain $g=2$ corresponds to $\cosh r= \sqrt{2}$ and $\sinh r= 1$, so we recover Eq. (\ref{eq-distribution-n}).
\par

\begin{figure}[t]
\centering
\includegraphics[trim = 0cm 0cm 0cm 0cm, clip, width=0.8 \linewidth, page=8]{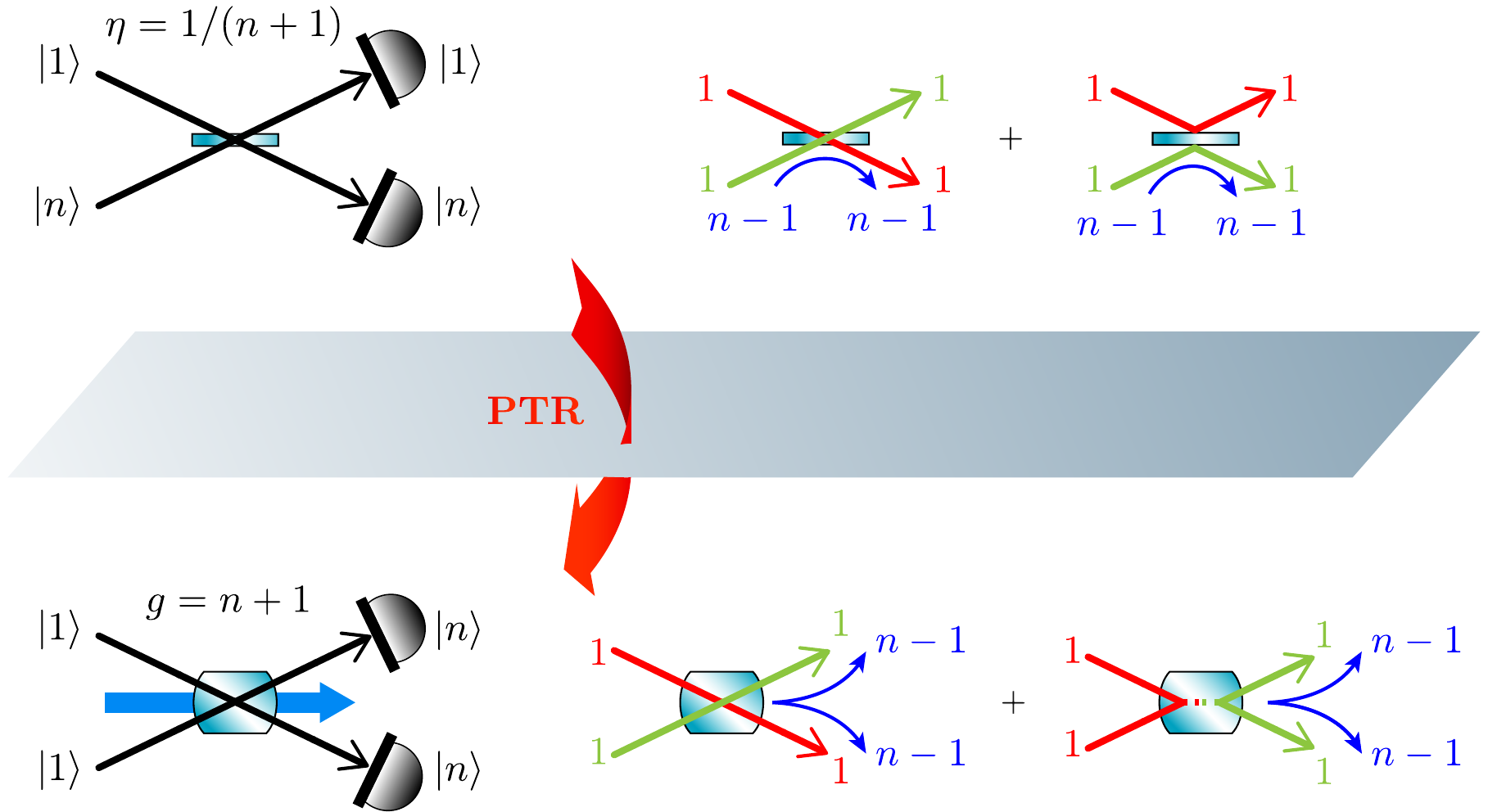}
\caption{Extended quantum interferometric suppression in an amplifier where the detection of $n$ photon pairs at the output is suppressed if the gain $g=n+1$. This is the PTR dual of the extended HOM effect when a single photon and $n$ photons impinge on the two input ports of an unbalanced beam splitter of transmittance $\eta=1/(n+1)$. The interference at play in the extended HOM is again the cancellation between a double reflection and double transmission (with $n-1$ other photons being always reflected from mode $\b$ to mode $\a$). Applying partial time reversal, this translates into the cancellation between the amplitude for the two photons crossing the crystal and the amplitude for the stimulated annihilation combined with stimulated emission of the two photons (accompanied, in both cases, with the stimulated emission of $n-1$ pairs). In other words, the input pair may be up-converted into the pump beam while $n$ pump photons are down-converted, or the input pair may simply be transmitted while $n-1$ pump photons are down-converted. Since the two scenarios are timelike indistinguishable, they interfere destructively.
}
\label{fig-extended-HOM}
\end{figure}

\subsection*{(F) Extension to a PDC with integer gain}
We may also consider the case where the gain takes a larger integer value, e.g., $g=3,\, 4,\cdots$. A closer look at Eq. (\ref{eq-output-coshr-sinhr}) reveals that the output term with $g-1$ photon pairs fully vanishes when $g$ is an integer. The corresponding distribution of the output photon pair number
\begin{equation}
\left| \bra{n,n} U^{\mathrm{PDC}}_{g} \ket{1,1} \right|^2 = \frac{(g-1)^{n-1}} {g^{n+2}} \,  (n+1-g)^2
\label{ref-extended-HOM-g}
\end{equation}
 is displayed in Fig. \ref{fig-output-state}b  for $g=3$ and 4.
This interferometric suppression $\bra{n,n} U^{\mathrm{PDC}}_{n+1} \ket{1,1}=0$, $\forall n$, can be interpreted as dual, under partial time reversal, to the extended HOM effect $\bra{n,1} U^{\mathrm{BS}}_{1/(n+1)} \ket{1,n}=0$ for a beam splitter of transmittance $1/(n+1)$, see Fig. \ref{fig-extended-HOM} (left). The latter effect is easy to understand as the interference between the amplitude with all $n+1$ photons being reflected and the amplitude with one photon of each mode being transmitted (the other $n-1$ being reflected). Indeed, the operator $\ad (\bd)^{n}$ transforms into
\begin{equation}
\frac{1 }{  (n+1)^{\frac{n+1}{2}}} \left( a'^\dagger - \sqrt{n} \, b'^\dagger \right)  \left( \sqrt{n} \, a'^\dagger +  b'^\dagger \right)^{n}   .
\end{equation}
The term proportional to  $  (a'^\dagger)^{n} \, b'^\dagger$ vanishes as a result of the cancellation of the term where the photon on mode $\a$ is reflected and the $n$ photons on mode $\b$ are reflected, together with the term where the photon on mode $\a$ is transmitted and one of the photons on mode $\b$ is transmitted (the other $n-1$ photons being reflected). This is sketched in  Fig. \ref{fig-extended-HOM} (right). More generally, the transition probability $1,n\to n,1 $ for a beam splitter of transmittance $\eta$ is given by
\begin{equation}
\left| \bra{n,1} U^{\mathrm{BS}}_{\eta} \ket{1,n} \right|^2 = (1-\eta)^{n-1} [ (n+1)\eta - 1]^2   ,
\label{ref-extended-HOM-eta}
\end{equation}
which is consistant with Eq. (\ref{ref-extended-HOM-g}) under partial time reversal, namely
\begin{equation}
\left| \bra{n,n} U^{\mathrm{PDC}}_{g} \ket{1,1} \right|^2 = \frac{1}{g}  \left| \bra{n,1} U^{\mathrm{BS}}_{1/g} \ket{1,n} \right|^2   .
\end{equation}
\par

\subsection*{(G) Experimental scheme}
\begin{figure}[t]
\centering
\includegraphics[trim = 0cm 0cm 0cm 0cm, clip, width=0.9\linewidth, page=5]{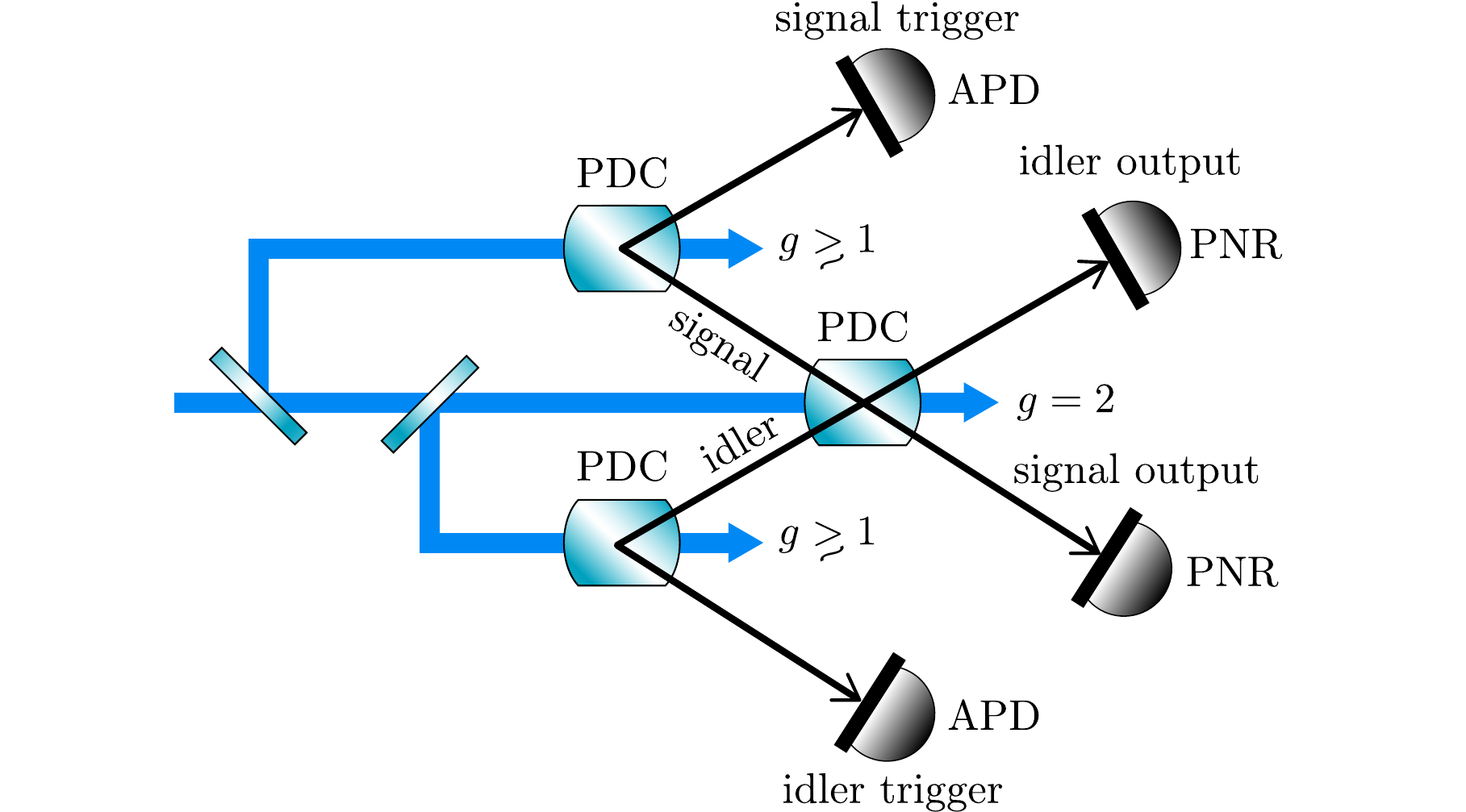}
\caption{Schematic of a potential demonstration of two-photon quantum interference in the amplification of light with gain~2. Two heralded single-photons sources (exploiting an avalanche photodetector APD) are used to feed the signal and idler modes of a PDC, and two photon-number resolving (PNR) detectors are used to monitor the presence of a single photon in each output port. The two-photon interference effect would be demonstrated by measuring a depletion of fourfold coincidences between the two trigger photons (heralding the preparation of two single photons) and the two output photons (here, the detectors should filter out the output states with $\ge 2$ photon on each mode). The dominant terms will then consist of the stimulated annihilation of the two input photons (witnessed by two trigger photons but no output photons) as well as the stimulated emission of a photon pair (witnessed by two output photons but no trigger photons). When the time lapse between the detection of the trigger and output photons is close to zero (which means a perfect match of the timing of the output photons originating from the input photons associated with the trigger photons) the two terms should interfere destructively.
}
\label{fig-experiment}
\end{figure}

The HOM effect is considered a most spectacular evidence of genuinely quantum two-boson interference, and we expect the same for its PTR counterpart as it admits no classical interpretation. The experimental verification of our effect can be envisioned with present technologies, as sketched in Fig.~\ref{fig-experiment}. We would need two single-photon sources, which could be heralded by the detection of a trigger photon at the output of a PDC with low gain (the single photon being prepared conditionally on the detection of the trigger photon in the twin beam). The two single photons would impinge on a PDC of gain 2, whose output modes should be monitored: the probability of detecting exactly one photon on each mode should be suppressed as a consequence of timelike indistinguishability. In principle, photon number resolution would be needed in order to discriminate the output term with one photon pair ($n$=1) from the terms with more pairs ($n$$\ge$2). The ability of  counting photons has become increasingly available over the last years (e.g., exploiting superconducting detectors), but this could also be achieved by splitting each of the two output modes into several modes followed by an array of \textsc{on/off} photodetectors.  Experiments involving PDC in three coherently pumped crystals have already been achieved recently \cite{HeuerPRL,HeuerPRA}, aiming at observing induced decoherence, so the proposed setup here should be implementable along the same lines.
The squeezing needed to reach a gain 2 amounts to 7.66~dB, which is high but lies in the range of experimentally accessible values (in the continuous-wave regime). The experiment could alternatively be carried out with a lower gain (especially in the pulsed regime) provided the observed dip is sufficient to rule out a classical interpretation. As a matter of fact, a coincidence probability lower than 1/4 would be needed (see Materials and Methods H), which could in principle be reached with a gain of 1.28 (i.e., a squeezing of 4.39~dB). Of course, the effect of losses should also be carefully analyzed in order to assess the feasibility of the scheme depicted in Fig.~\ref{fig-experiment}.
\par

Demonstrating this effect would be invaluable in view of the fact that the HOM dip is widely used to test the indistinguishability of single photons and to benchmark mode matching: it witnesses the fact that the photons are truly indistinguishable (they admit the same polarization and couple to the same spatio-temporal mode). For example, HOM  experiments have been used to test the indistinguishability of single photons emitted by a semiconductor quantum dot in a microcavity \cite{Yamamoto}, while the interference of two single photons emitted by two independently trapped rubidium-87 atoms has been used as an evidence of their indistinguishability \cite{Grangier}. The HOM effect has also been generalized to 3-photon interference in a 3-mode optical mixer \cite{Campos}, while the case of many photons in two modes has been analyzed in \cite{Nakazato}, implying a possible application of the quantum Kravchuk-Fourier transform \cite{Stobinska}. We anticipate that most of these ideas could extend to interferences in an active optical medium.

\subsection*{(H) Classical baseline}
The two-photon quantum interference effect in amplification cannot be interpreted within a classical model of PDC, where a pair can be annihilated or created with some probability.  We have two possible indistinguishable paths (the photon pair either going through the cristal or being replaced by another one) with equal individual probabilities but opposite probability amplitudes, hence the resulting probability vanishes (whereas the two probabilities would add for classical particles).  In order to assess an experimental verification of this effect, it is necessary to establish a classical baseline, namely to determine the depletion of the probability of coincident detections that can be interpreted classically. As a guide, consider first a classical model of the HOM effect where the two input photons are distinguishable. We have to add the double-transmission probability $|A_\mathrm{dt}|^2$ with the double-reflection probability $|A_\mathrm{dr}|^2$ since these two paths can be distinguished. Then, the classical probability for coincident detections is
\begin{equation}
P_\mathrm{cl} (\eta)= |A_\mathrm{dt}|^2 + |A_\mathrm{dr}|^2 = \eta^2 + (1-\eta)^2   ,
\end{equation}
to be compared with $P_\mathrm{coinc} (\eta)$ of Eq. (\ref{eq-Pcoinc-BS}). For a 50:50 beam splitter, $P_\mathrm{cl} (1/2)= 1/2$, hence a depletion below 50\% ensures that that the dip is quantum. Similarly, in a classical model of PDC, we can distinguish the path where the two input photons are transmitted (probability $|A'_\mathrm{dt}|^2$) from the path where they are replaced by another pair (probability $|A'_\mathrm{st}|^2$). Thus, the classical probability for coincident detections is
\begin{equation}
P'_\mathrm{cl} (g)= |A'_\mathrm{dt}|^2 + |A'_\mathrm{st}|^2 = \frac{1+(g-1)^2 }{ g^3}    ,
\end{equation}
to be compared with  $P'_\mathrm{coinc} (g)$ of Eq. (\ref{eq-Pcoinc-PDC}). For a gain-2 PDC, $P'_\mathrm{cl} (2)=1/4$, hence we need to have a coincidence probability less than 25\% in order to exclude a classical interpretation of the dip. Interestingly, this suggests that a gain lower than 2 may be utilized for demonstrating this quantum effect, thus lowering the experimental requirements. Solving $P'_\mathrm{coinc} (g) = 1/4$, we obtain $g = 1.28$, implying that the partial indistinguishability between the two paths achieved with this lower gain would be sufficient for observing a quantum effect. With this gain, the probability amplitude corresponding to the transmission through the cristal is larger than (minus) the one corresponding to double stimulated events, but the partial destructive interference between the two paths is sufficient to reduce the coincidence probability to 1/4.
\par


\medskip
\noindent{\textbf{Acknowledgements} \\
We thank Ulrik L. Andersen, Maria V. Chekhova, Claude Fabre, Virginia D'Auria, Linran Fan, Radim Filip, Saikat Guha, Dmitri Horoshko, Mikhail I. Kolobov, Julien Laurat, Klaus M\"olmer, Romain Mueller, Ognyan Oreshkov, Olivier Pfister, Wolfgang P. Schleich, S\'ebastien Tanzilli as well as an anonymous referee for useful comments. 
M.G.J. acknowledges support from the Wiener-Anspach Foundation. This work was supported by the Fonds de la Recherche Scientifique -- FNRS under project PDR T.0224.18.\\

\noindent\textbf{Author contributions} \\
N.J.C. designed research; and N.J.C. and M.G.J. performed research, derived the formulas, discussed the results, and wrote the paper.\\

\noindent\textbf{Competing financial interests} \\ 
The authors declare no competing financial interests.\\

\noindent\textbf{Additional information} \\
Correspondence should be addressed to N.J.C. \\
Email: ncerf@ulb.ac.be

\bibliography{PNAS-archive-version}

\end{document}